\documentclass[showpacs,twocolumn,pre,floatfix]{revtex4}
\usepackage{psfrag,epsfig,amsfonts,amssymb,amsmath,wasysym}
\usepackage{dcolumn}

\usepackage[normalem]{ulem}
\usepackage{color}

\begin{document}
\newcommand{\RR}{{\mathbb R}}
\newcommand{\pos}{x}
\newcommand{\cm}{X}
\newcommand{\ori}{\phi}
\newcommand{\poscm}{y}
\newcommand{\per}{L}
\newcommand{\kB}{k}
\newcommand{\fluc}{\zeta}

\title{Exploiting lattice potentials for sorting chiral particles}

\author{David Speer$^{1}$}
\author{Ralf Eichhorn$^{2}$}
\author{Peter Reimann$^{1}$}
\affiliation{$^{1}$Universit\"at Bielefeld, Fakult\"at f\"ur Physik, 33615 Bielefeld, Germany
\\
$^{2}$NORDITA, Roslagstullsbacken 23, 10691 Stockholm, Sweden}

\begin{abstract}
Several ways are demonstrated of how periodic 
potentials can be exploited for sorting
molecules or other small objects which 
only differ by their chirality.
With the help of a static bias force, the
two chiral partners can be made to move
along orthogonal directions.
Time-periodic external forces even lead to motion
into exactly opposite directions.
\end{abstract}


\maketitle

Chiral particles are extended objects which are 
non-superposable with their mirror image.
So-called enantiomers, i.e. chemically identical molecular 
species with opposite chirality 
play a crucial role in Chemistry, Biology, and Medicine
due to the omnipresence of chiral molecules in living 
organisms 
but with only one of the two chiral partners actually 
being present.
Accordingly, enantiomers in drugs, pesticides etc. 
have very different effects on an organism and thus 
their separation is of great importance.
Established methods of separating enantiomers 
mostly exploit some kind of chiral 
selector \cite{ahu97},
i.e. some materials, structures, 
or ancillary molecules
which themselves exhibit an intrinsic 
chirality.
Their main disadvantage is that essentially
every enantiomer species requires a 
different selector.
Therefore, several alternative 
concepts have recently been put forward.
A first promising direction proposes to utilize
appropriate microfluidic flows, 
such as vortices
\cite{kos06} or shear flows 
\cite{mar09,wat09}.
Second, photoinduced separation 
by means of suitably chosen electromagnetic 
fields has been theoretically predicted
in Ref. \cite{spi09}.
A third approach to exploit a structure without an intrinsic 
chirality is due to de Gennes \cite{gen99}, predicting 
qualitatively that, according to Curie's principle \cite{cur94},
small chiral crystalls should slide down an inclined plane
along directions which slightly differ for the two chiral 
partners, provided thermal noise 
is negligible.
Here,
we further pursue this 
approach, 
showing that with the help of periodic 
potentials the two chiral partners even can be made to
move into opposite directions,
with remarkable persistence 
against thermal noise.

Apart from ``true'' (bio-) molecular enantiomers,
we also have in mind chiral 
nano- and micro-particles, 
e.g. helically shaped 
nonmotile bacteria \cite{mar09}
and artificial flagellae \cite{zha09}, 
carbon nanotubes,
chiral colloidal clusters \cite{zer08},
or  ferromagnetic nano-propellers \cite{goh09}.
The periodic potentials we are proposing 
to utilize for sorting those chiral particles 
may be realized e.g. by means of 
crystal surfaces \cite{mir05},
optical lattices \cite{opt},
periodic micro- and nano-structures \cite{str},
or magnetic bubble lattices \cite{mag}.

\begin{figure}
\epsfxsize=0.9\columnwidth
\epsfbox{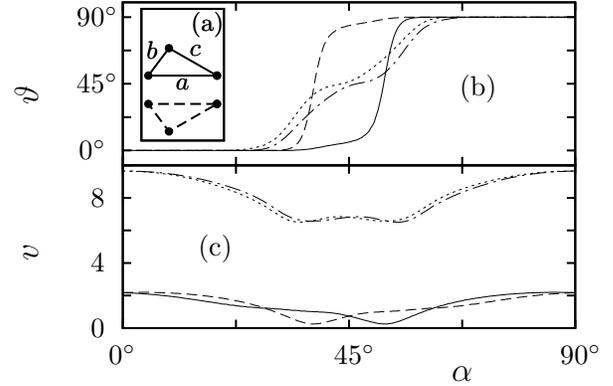} 
\caption{(a) Solid: Triangle, specified by 
$a$, $b$, $c$.
Dashed: Its chiral partner.
(b) and (c): Direction $\vartheta$ and modulus $v$ of the net velocity
$\vec v=\vec e_\vartheta v$ versus the direction $\alpha$ of a static
bias $A$ in (\ref{5}) by numerically solving (\ref{2})-(\ref{6})
with parameters as specified below (\ref{6}).
Solid: $\kB T=0.08$, $A=3.6$, $a=0.4$, $b=0.23$, $c=0.34$
(proportional to the solid triangle in (a)).
Dashed: Same but for the chiral partner
(dashed in (a)).
Dashed-dotted and dotted: Same but for $A=10$.
}
\label{fig1}
\end{figure}

Most of us are not very accustomed to think
in terms of chiral symmetry and symmetry-breaking,
especially in combination with the crystal
symmetries of a
periodic potential.
For this reason only, we mainly focus on
the simplest possible setup \cite{kos06},
namely the two-dimensional
dynamics in a square lattice potential
of a ``minimal'' planar, chiral
``molecule'',
consisting of three identical, rigidly 
coupled ``atoms'' or other small 
objects with broken mirror symmetry,
see Fig. 1a.
All basic effects and mechanisms are 
recovered in three dimensions and also 
for more general lattices, 
but are much more cumbersome 
to visualize and explain.
This very general validity of our main 
results will be exemplified for various 
other chiral ``molecules'' in the end.
We thus consider the
two-dimensional dynamics 
\begin{equation}
m_i \ddot{\vec\pos}_i(t) = 
-\gamma_i \dot{\vec\pos}_i(t) + {\vec F}(\vec{\pos}_i(t),t)+{\vec f}_i+{\vec \xi}_i(t) \ .
\label{1}
\end{equation}
Dots indicate time derivatives,
$\vec{\pos}_i={\vec e}_1 \pos_{i,1}+{\vec e}_2 \pos_{i,2}$ are
the ``atom positions''
($i=1,2,...,N$) 
in Cartesian coordinates ${\vec e}_\nu$ ($\nu=1,2)$, 
$m_i$ their mass, and $\gamma_i$ their dissipation coefficient, 
e.g.\ due to an ambient fluid.
In particular, for the triangular particles (Fig. 1a) we
have $N=3$ and $i$-independent $\gamma_i$ and $m_i$.
The force field ${\vec F}(\vec{\pos},t)$ 
is partly due to a ``lattice potential''
(see 
below) 
and partly due to an externally applied driving, 
typically via electrophoresis.
Under these conditions, hydrodynamic interactions 
are screened \cite{lon96} and therefore safely 
negligible \cite{kos06,mar09}.
The internal constraining forces, maintaining e.g. 
the triangular shape in Fig. 1a,
are represented by ${\vec f}_i$,
and thermal fluctuations are modelled as usual 
by unbiased Gaussian white noise 
${\vec \xi}_i(t)={\vec e}_1 \xi_{i,1}(t)+{\vec e}_2 \xi_{i,2}(t)$,
satisfying the fluctuation dissipation relation
$\langle\xi_{i,\mu}(s)\xi_{j,\nu}(t)\rangle=2\gamma_i \kB T\delta_{ij}\delta_{\mu\nu}\delta(s-t)$
with $T$ the ambient temperature and $\kB$ Boltzmann's constant.
The position of the rigid ``molecule'' is
conveniently specified by the so-called
center of friction \cite{kos06}
$\vec{\cm}:=\sum_{i=1}^N \gamma_i\vec{\pos}_i/\sum_{i=1}^N\gamma_i$,
and its orientation by the angle $\ori$
between the $\vec e_1$ axis and
the position of ``atom 1'' relative to the
center of friction: $\ori := \varangle (\vec e_1,\vec x_1 -\vec X)$.
Rewriting (\ref{1}) in terms of $\vec{\cm}$ and $\ori$ 
to get rid of the constraining forces $\vec f_i$
is a basic mechanics exercise.
Further, for the 
very small objects we 
have in mind, inertia effects are negligible 
\cite{rei02,static},
yielding \cite{kos06}
\begin{eqnarray}
\dot{\vec \cm}(t) & = & \frac{\sum_{i=1}^N {\vec F}(\vec{\pos}_i(t),t)}{\sum_{i=1}^N \gamma_i}+\vec\fluc (t) \, ,
\label{2}
\\
\dot{\ori} (t)& = & 
\frac{{\vec e}_3 \cdot \sum_{i=1}^N \vec\poscm_i(t)\times {\vec F}(\vec{\pos}_i(t),t)}{\sum_{i=1}^N \gamma_i\, \poscm_i^2}+\fluc_\varphi(t) \, ,
\label{3}
\\
\vec{\pos}_i(t) & = & {\vec\cm}(t)+\vec{\poscm}_i(t)\ , \ \ 
\vec{\poscm}_i(t) = {\bf O}(\ori(t)) \vec{\poscm}_i(0) \, .
\label{4}
\end{eqnarray}
In (\ref{3}), vectors are temporally embedded into 
$\RR^3$ with standard scalar and vector products
$\cdot$ and $\times$.
In (\ref{4}), ${\bf O}(\ori)$ is a rotation matrix
with elements $O_{11}=O_{22}=\cos\ori$
and $O_{12}=-O_{21}=-\sin\ori$.
Thus, $\vec\poscm_i(t)$ are 
the particle postions relative to the center of friction
with convention $\ori(0)=0$ and with 
$t$-independent modulus $\poscm_i:=|\vec\poscm_i(t)|$.
Finally, $\vec\fluc(t)$
and $\fluc_\ori(t)$ are independent Gaussian white noises
with $\langle\fluc_\mu(s)\fluc_\nu(t)\rangle=2\kB T\delta_{\mu\nu}\delta(s-t)/\sum\gamma_i$
and $\langle\fluc_\ori(s)\fluc_\ori(t)\rangle
=2\kB T\delta(s-t)/\sum \gamma_i  \poscm_i^2$.

As already said, the force field consists of two parts,
\begin{equation}
\vec F(\vec{\pos},t) =  {\vec e}_\alpha \, A(t) -\vec{\nabla} U(\vec{\pos}) \ ,
\label{5}
\end{equation}
namely a spatially homogeneous, externally
applied force along the direction
${\vec e}_\alpha:={\vec e}_1\cos\alpha + {\vec e}_2\sin\alpha$
and a Gaussian square lattice potential with period $\per$:
\begin{equation}
U(\vec{\pos})=u \sum_{m,n=-\infty}^\infty 
\exp\{-\frac{(\vec{\pos}- [m \vec e_1 + n \vec e_2] \per )^2}{2\sigma^2}\} \ .
\label{6}
\end{equation}
For this potential with $u>0$ and $u<0$ as well as for
various other potentials 
we always found similar results.
Focusing on $u>0$ from now on, the natural energy scale is
the potential barrier 
$\Delta U:=U(\vec e_1 \per/2)-U(\vec 0)$ separating adjacent
potential wells.
We henceforth adopt time, energy, and length units so that 
$\min_i \gamma_i=1$, $\Delta U=1$, $\per=1$, 
and 
focus on 
$\sigma=L/4$ \cite{f1}.

The quantity of central interest is the net velocity 
$\vec v=\vec e_\vartheta v$,
obtained by averaging $\dot{\vec\cm}(t)$ over time.
Obviously, rotating the force field (\ref{5}) by $90^\circ$
leaves the potential (\ref{6}) invariant and entails 
a rotation of $\vec v$ by $90^\circ$.
Hence, it is sufficient to focus on $\alpha \in [0^\circ,90^\circ$].
Likewise, one readily sees that
$A(t) \mapsto -A(t)$ implies $\vec v\mapsto-\vec v$.

We first consider $t$-independent $A$, i.e. the force
field (\ref{5}) derives from a tilted periodic potential.
For $A=0$ symmetry implies $\vec v=\vec 0$.
For $A\not =0$ the salient point is to realize
that there exists no symmetry argument why 
two ``molecules'' of opposite chirality 
should travel down the tilted periodic 
potential with identical velocities $\vec v$.
Following de Gennes \cite{gen99}, we thus can invoke 
Curie's principle to conclude \cite{cur94,rei02}
that generically 
(i.e. up to parameter sets of measure zero) 
the velocities will indeed be different.
In other words,
(practically) {\em any tilted periodic 
potential can separate chiral partners} 
via their velocities. 
The main remaining problem pinpointed by De Gennes
is the quantitative efficiency of the effect.

Fig. 1 provides those quantitative details 
in a typical case.
We see that the velocities $\vec v$ 
of the two chiral partners
are indeed disappointingly similar, 
except around $\alpha=45^\circ$.
The explanation 
is as follows:
For small thermal energies $\kB T$ and small
bias $A$, the particles travel extremely slowly
by thermally activated hopping from one
local minimum of the tilted periodic potential
to the next.
For any given orientation $\alpha$ there exists a critical 
tilt $A$ in (\ref{5}) at which
certain local minima
disappear by annihilation (collision) with 
saddle points, giving rise to ``running solutions''.
For $\kB T=0$ (deterministic limit), 
these solutions travel either parallel 
to $\vec e_1$ or to $\vec e_2$, 
and for small $\kB T>0$ 
still almost so.
Roughly speaking, the direction ``closer''
to that of the static bias $\vec e_\alpha A$
is preferred, but due to the broken mirror
symmetry, the direction actually switches already at some
$\alpha < 45^\circ$ for one chiral partner
and symmetrically at $\alpha > 45^\circ$
for the other (solid and dashed in Fig. 1).
Since these considerations do not depend
on any details of the model we can conclude that
{\em a separation by (almost) 
$90^\circ$ is generic
for $\alpha=45^\circ$, small $\kB T$, 
and $A$ close to criticality}.
Upon further increasing $A$, the
deterministic running solutions speed up and 
bifurcate into new ones, 
``locked'' \cite{opt,static} along directions of the form 
$n \vec e_1+m\vec e_2$ with increasingly 
large 
integers 
$n$ and $m$
(dashed-dotted and dotted in Fig. 1)
and with 
$\vec v\to \vec e_\alpha A N/\sum\gamma_i$ 
for $A\to\infty$.
Likewise, for finite $\kB T$ the
deterministically ``sharp'' bifurcations
get washed out (Fig. 1) and 
$\vec v\to \vec e_\alpha A N/\sum\gamma_i$ 
for $\kB T\to\infty$.

\begin{figure}
\epsfxsize=0.95\columnwidth
\epsfbox{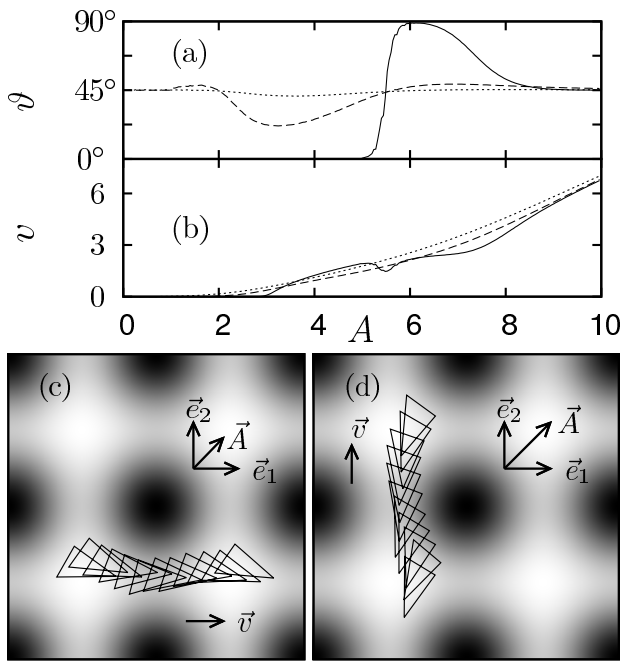} 
\caption{(a) and (b): 
Direction $\vartheta$ and modulus $v$ of the net
velocity $\vec v=\vec e_\vartheta v$ for the solid
triangle from Fig.\ 1a versus static
bias $A$ with fixed direction $\alpha=45^\circ$.
Shown are numerical solutions of
(\ref{2})-(\ref{6})
for $\kB T=0.014$ (solid),
$\kB T=0.16$ (dashed),
and $\kB T=0.32$ (dotted).
Other parameters as in Fig. 1.
(c) and (d): Illustration of the triangle's
motion for $A= 4$ (c)
and $A= 6$ (d) at $\kB T=0.014$.
The static bias is indicated by 
$\vec A:=\vec e_\alpha A$
and the periodic potential (\ref{6}) 
as ``shaded background''.
The triangle motion is shown
for a time-span
of about 14.5 time-units in (c) and about 8.5 in (d),
and then continues periodically up to 
noise effects (not shown).
}
\label{fig2}
\end{figure}

Thus focusing on $\alpha=45^\circ$,
the dependence of the velocity $\vec v$
on the bias $A$ is shown in Fig. 2 (a,b).
For symmetry reasons, 
{\em the velocities of the two chiral
partners are now equal in modulus and symmetric
about $\alpha=45^\circ$} (see also Fig. 1).
Remarkably enough,
for some $A$-values, one triangle
moves (practically) parallel to $\vec e_1$
(Fig. 2c) 
and thus its chiral partner 
parallel to $\vec e_2$
(not shown in Fig. 2),
while for some different $A$-values 
it is exactly the other way round (Fig. 2d).
In other words,
{\em one and the same triangle may move along orthogonal 
directions for two different $A$-values}.

Turning to periodic $A(t)$ in (\ref{5}),
our so far findings quite naturally
suggest the following idea:
We select $\alpha=45^\circ$ and two static 
bias values $A_1$ and $A_2$
with velocities $\vec v_1=v_1 \vec e_1$ and 
$\vec v_2=v_2\vec e_2$
(e.g. $A_1=4$ and $A_2=6$ for the solid lines 
($\kB T=0.014$) in Fig. 2a,b)
and exploit that the signs of $v_1$ and $v_2$ 
can be arbitrarily chosen by adjusting 
the signs of $A_1$ and $A_2$
(recall that $A \mapsto -A$ implies $\vec v\mapsto-\vec v$).
If we now construct a time-periodic $A(t)$ 
which takes the value $A_1$ during 
a fraction $p\in[0,1]$ of its total period $\tau$
and the value $A_2$ during the rest of 
the period, the resulting time 
averaged velocity 
will be 
$\vec v=p\vec v_1+(1-p)\vec v_2$,
provided $\tau$ is so large that transient
effects after each jump of $A(t)$ are negligible.
Therefore, the molecule can be steered into
\emph{any} direction on the two-dimensional
plane by varying $p$ and adapting the signs of
$A_{1,2}$.
In particular, we will encounter a situation
where $\vec v$ is orthogonal to the force direction 
$\vec e_\alpha$.
E.g. from the solid lines in Fig.\ 2 we can read off
that the 
triangle 
will move with such a velocity $\vec v\perp \vec e_\alpha$
if we choose $A_1=-4$, $A_2=6$, and $p\approx 2/3$ 
to account for the difference in modulus of 
the corresponding velocities
$\vec v_1\approx -1.2\, \vec e_1$ and
$\vec v_2\approx 2.4\, \vec e_2$.
The net velocity $\vec v$ of the chiral partner follows
from the above mentioned symmetry about
$\alpha =45^\circ$:
This symmetry applies to both $\vec v_1$
and $\vec v_2$ separately, and hence 
also to $\vec v=p\vec v_1+(1-p)\vec v_2$.
Altogether, {\em the two chiral partners 
can thus be forced to move into exactly 
opposite directions}.
Deviations due to the so far neglected transient 
effects after each jump of $A(t)$ are -- at least
for not too small $\tau$-values -- small and thus
can be compensated by adjusting
$p$ and/or $A_{1,2}$.

\begin{figure}
\epsfxsize=0.95\columnwidth
\epsfbox{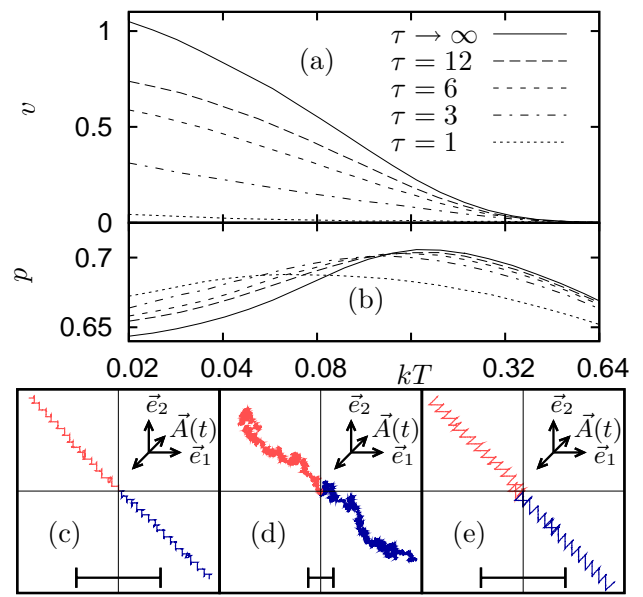} 
\caption{(a): 
Absolute velocity $v$
versus thermal energy $\kB T$ (logarithmic scale) 
for the same system as in Fig. 1 
but with a $\tau$-periodic driving $\vec e_\alpha A(t)$ 
with $\alpha=45^\circ$ and $A(t)$
taking the value $A_1=-4$ during a fraction $p$ 
of the period $\tau$ and the value $A_2=6$ 
during the rest of the period.
(b):
The corresponding $p$-values, adjusted as described 
in the main text so that the two chiral partners 
move into exactly opposite directions.
(c) Typical single-particle trajectories $\vec\cm (t)$
for the two chiral partners (light red and dark blue) 
with 
$t\in[0,\, 100]$,
$\vec\cm(0)=\vec 0$, 
$\kB T=0.02$, $p=0.66$, and $\tau = 6$.
Other parameters as in (a).
The bar indicates 50 lattice periods
and the double arrow the periodic driving \cite{f1}.
(d) Same but for a much larger thermal energy 
$\kB T=0.32$ and $t\in[0,\, 6000]$, $p=0.69$.
(e) Same as in (c) but for a
very different triangel with $a=2.5$, $b=2.2$, $c=1.1$ 
(cf. Fig. 1), and a driving with 
$A_1=-7$, $A_2=14$, $p=0.8$ \cite{f1}.
}
\label{fig3}
\end{figure}

Fig. 3 shows that these ideas indeed 
work out in practice, 
and  in fact down to surprisingly small 
time-periods $\tau$ and up to 
remarkably large thermal energies $\kB T$.
Note that while the velocities in Fig. 3a
are long-time averages, Figs. 3c-e exemplify 
single-particle trajectories of moderate 
duration.
Hence the thermal noise still leads to quite notable 
random fluctuations of each trajectory $\vec X(t)$ 
around the average behavior, especially in Fig. 3d.
Only in Figs. 3c,e we still can see the expected 
``steps'' of $\vec X(t)$ at jumps of $A(t)$.

Our above recipe for tailoring transport directions
can be readily extended
to arbitrary velocities $\vec v_1$ and $\vec v_2$,
provided they are not parallel to each other:
Then, as before, $\vec v=p\vec v_1+(1-p)\vec v_2$
can be made to point along any direction by
properly choosing $p$ and the signs of $A_{1,2}$.
Intuitively and in view of Fig. 2,
it is quite clear that generically one will always
be able to find two bias values $A_1$ and $A_2$ 
with non-parallel
velocities $\vec v_1$ and $\vec v_2$.
We thus can conclude that
{\em chiral partners can (practically) always be made 
to move into opposite directions by means 
of a suitably tailored periodic driving force}.
Fig. 3e exemplifies this generalized 
theoretical scheme for comparatively ``large''
triangular particles
and Fig. 4 for a representative selection of
more general chiral ``molecules''.
Generalizations involving more than two ``static velocities''
$\vec v_i$ and the concomitant optimization problems
point into interesting directions for future research.

\begin{figure}
\epsfxsize=0.95\columnwidth
\epsfbox{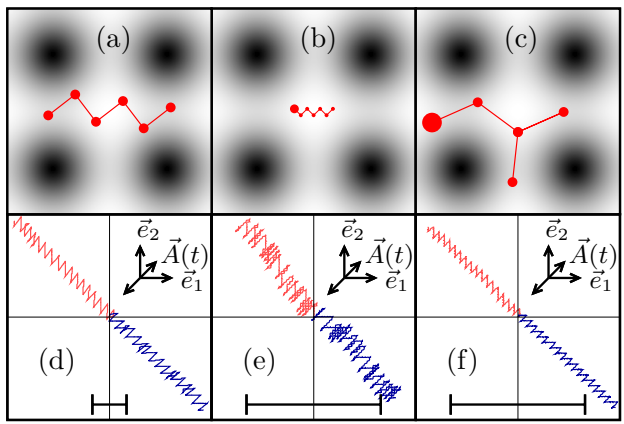} 
\caption{(a)-(c): Further examples of chiral 
``molecules''. Dots indicate the
constituting ``atoms'', lines their rigid 
coupling, and the ``shaded background''
the periodic potential (\ref{6}).
Adopting units as specified below (\ref{6}),
$\gamma_i=1$ for all ``atoms'' except for
the larger dots in b and c,
representing $\gamma_i=3$.
(d) Typical single-particle trajectories $\vec\cm (t)$
for the molecule from (a) and 
its chiral partner (light red and dark blue),
obtained by numerically simulating
(\ref{2})-(\ref{6}) with $t\in[0,\, 100]$, 
$\vec\cm(0)=\vec 0$, and $\kB T=0.02$.
Parameters of the periodic driving (see main text):
$\alpha=45^\circ$, $A_1=7.8$, $A_2=-11.7$,
$\tau = 6$, $p=0.71$.
The bar indicates 50 lattice periods
and the double arrow the periodic driving.
(e): Same but for the ``molecule'' from (b)
and $A_1=7$, $A_2=-10.5$, $p=0.78$ \cite{f1}.
(f): Same but for the ``molecule'' from (c)
and $A_1=3$, $A_2=-6$, $p=0.85$ \cite{f1}.
}
\label{fig4}
\end{figure}

In conclusion, periodic potentials can act as 
very effective and versatile selectors for
sorting small objects which only differ by their 
chirality.
Static bias forces make the two 
chiral partners move into directions which differ
by up to $90^\circ$ (Figs. 1,2).
Appropriately chosen time-periodic forces even lead to
motion into exactly opposite directions (Figs. 3,4).
A major advantage compared to many other 
separation concepts \cite{ahu97} is that
one and the same periodic potential may act
as an efficient selector for quite different 
chiral particle species by suitably adapting 
the time-periodic driving force.
Furthermore, the separation mechanisms 
are remarkably robust against thermal 
noise.
The basic symmetry breaking conditions at the
origin of all these effects are generically 
satisfied for much more general systems than 
in (\ref{1})-(\ref{6}), including 
three spatial dimensions,
finite inertia effects, 
other chiral objects
and crystal potentials.
An experimental proof of principle
for chiral micro-particles \cite{zha09,zer08,goh09}
moving in a periodically structured microfluidic 
device \cite{str} is presently under construction
in the Anselmetti lab at Bielefeld University.

\begin{center}
\vspace{-5mm}
---------------------------
\vspace{-4mm}
\end{center}
This work was supported by Deutsche Forschungsgemeinschaft 
under SFB 613 and RE1344/5-1

\end{document}